\documentclass[aps,superscriptaddress,showpacs,floatfix]{revtex4-1}

\usepackage{hyperref}
\usepackage{color}
\usepackage{graphicx}	% Include figure filed

\begin{document}

\title{String Theory Based Predictions for Nonhydrodynamic Collective Modes in Strongly Interacting Fermi Gases}

\author{H.~Bantilan}
\affiliation{School of Mathematical Sciences, Queen Mary University of London, E1 4NS, UK}
\affiliation{Centre for Research in String Theory, School of Physics and Astronomy, Queen Mary University of London, E1 4NS, UK}
\affiliation{DAMTP, University of Cambridge, CB3 0WA, UK}
\author{J.T.~Brewer}
\affiliation{Center for Theoretical Physics, MIT, Cambridge, MA 02139, USA}
\author{T.~Ishii}
\affiliation{Department of Physics, 390 UCB, University of Colorado at Boulder, Boulder, Colorado 80309, USA}
\affiliation{Center for Theory of Quantum Matter, University of Colorado, Boulder, Colorado 80309, USA}
\author{W.E.~Lewis}
\affiliation{Department of Physics, 390 UCB, University of Colorado at Boulder, Boulder, Colorado 80309, USA}
\author{P.~Romatschke}
\affiliation{Department of Physics, 390 UCB, University of Colorado at Boulder, Boulder, Colorado 80309, USA}
\affiliation{Center for Theory of Quantum Matter, University of Colorado, Boulder, Colorado 80309, USA}

\date{\today}

\begin{abstract}
Very different strongly interacting quantum systems such as Fermi gases, quark-gluon plasmas formed in high energy ion collisions and black holes studied theoretically in string theory are known to exhibit quantitatively similar damping of hydrodynamic modes. It is not known if such similarities extend beyond the hydrodynamic limit. Do non-hydrodynamic collective modes in Fermi gases with strong interactions also match those from string theory calculations? In order to answer this question, we use calculations based on string theory to make predictions for novel types of modes outside the hydrodynamic regime in trapped Fermi gases. These predictions are amenable to direct testing with current state-of-the-art cold atom experiments.
\end{abstract}

\maketitle

\section{Introduction}

Traditional descriptions of quantum matter rely heavily on approaches formulated in terms of particle-like constituents, ranging from ordinary electrons, nuclei and photons to fermions in Fermi liquids and phonons in superfluids. These particle-based descriptions are and have been extremely successful and provide the backbone of modern physics' ability to describe nature.

Recently, however, it  has become clear that there is an ever growing class of strongly interacting systems where particle-based descriptions simply do not work, such as in the mysterious normal state of high temperature superconductors, fractional quantum hall effect, extremely hot quark gluon plasmas and ultracold quantum gases. Precision experiments in the past decade suggest that despite the fact that quark-gluon plasmas and ultracold quantum gases differ by no less than 18 orders of magnitude in temperature, their ability to flow around obstacles  is very similar. This notion can be quantified by comparing the ratio of shear viscosity $\eta$ to the entropy density $s$ for both normal fluids, finding $\eta k_B/\hbar s\simeq 0.2\pm0.1$ for the quark-gluon plasma \cite{Gale:2012rq} and $\eta k_B/\hbar s\simeq 0.2-0.4\pm0.1$ for ultracold quantum gases \cite{Cao:2010wa}, where $\hbar$ is Planck's constant divided by $2\pi$ and $k_B$ is Boltzmann's constant. It should be noted that in both these systems the precise determination of $\eta k_B/\hbar s$ is still an ongoing effort and requires further experimental and theoretical studies \cite{PhysRevA.75.043612,2011AnPhy.326..770E,PhysRevLett.109.020406,2015PhRvL.115b0401J,Bluhm:2015raa}. Nevertheless, it seems to be an established fact that the quark gluon plasma and ultracold quantum gases have (minimum) values of $\eta k_B/\hbar s$ which differ only by a factor of a few at most. 

Unlike ordinary liquids which have particle-like constituents, long-lived particles do not seem to exist for strongly interacting quantum liquids such as hot quark gluon plasmas and ultracold quantum gases. This breakdown of traditional particle-based methods motivates the search for new, non-particle based descriptions of strongly interacting quantum matter. One recently developed theoretical tool for strongly interacting systems that does not rely on any particle-based description is the conjectured duality between  classical black holes and strongly interacting quantum field theories originating in string theory \cite{Maldacena:1997re}. Within this framework, it is possible to calculate the friction coefficient $\eta/s$ for a strongly interacting quantum liquid. One obtains $\eta k_B/\hbar s=1/4\pi\simeq 0.08$, which is close to the experimentally determined values for both hot quark gluon plasmas and ultracold quantum gases \cite{Policastro:2001yc}.

The fact that very different systems such as hot quark gluon plasmas, ultracold quantum gases and black holes in low energy string theory have quantitatively similar values of $\eta/s$ has led to the conjecture that transport properties in strongly interacting quantum liquids are approximately universal \cite{Brewer:2015ipa}. In cold Fermi gases, the presence of a quantum critical point at zero density and temperature also gives rise to universal scaling properties of thermodynamic and transport quantities, cf. Refs.~\cite{Nikolic:2007zz,PhysRevA.86.013616}. This quantum critical point universality is different from the universality discussed above, which is thought to originate from the strongly coupled nature of different systems. Indeed, neither quark gluon plasmas nor the black hole considered here are close to a quantum critical point. Universality would imply that the transport properties of a given strongly interacting liquid are matched quantitatively by any other realization of a strongly interacting liquid as long as basic symmetry requirements are fulfilled, even if the liquids themselves have very different constituents, temperatures or densities. In the case at hand, exact universality cannot be expected, as is already evident from the somewhat different values of $\eta/s$ for quark gluon plasmas, ultracold quantum gases and black holes. However, studying approximate symmetries in physics has had enormous successes in the past, as is evident in the cases of chiral symmetry of the strong force and parity symmetry of the weak force to name two prominent examples. Thus, it may nevertheless be useful to consider the possibility of approximate universality concerning the transport properties of different strongly interacting quantum liquids.

The aim of the present work is to test approximate transport universality of strongly interacting quantum fluids by making experimentally testable predictions. Examining the case of black holes as one example of strongly interacting quantum fluids, one finds that a robust transport feature of black holes is the existence of so-called non-hydrodynamic quasi-normal modes, which characterize the ring-down of the black hole after some perturbation occurred (see Ref.~\cite{Berti:2009kk} for an in-depth review of the physics of quasinormal modes of black holes). If transport properties of strongly interacting fluids are approximately universal, we expect these non-hydrodynamic modes to be realized in ultracold quantum gases. In the remainder of this article, we assume that a commonly studied experimental setup for ultracold Fermi gases admits a gravity-dual approximation in terms of a black hole. We then proceed to make quantitative predictions for the frequencies and damping rates of non-hydrodynamic modes based on this model. These predictions are within the reach of current state-of-the-art experiments, and can be checked by higher-precision experimental data. A detection of non-hydrodynamic modes in ultracold Fermi gases would support the conjectured approximate universality of transport properties among strongly interacting quantum fluids

The remainder of the work is organized as follows: in section \ref{sec:osc}, we review the experimental setup to study collective oscillations in ultracold Fermi gases. In section \ref{sec:string}, the theoretical calculations of the non-hydrodynamic modes are described, including a summary of assumptions used to convert these calculations to experimentally accessible predictions. Section \ref{sec:res} contains our results and we conclude in section \ref{sec:conc}.

\section{Collective Oscillations in Ultracold Fermi Gases}
\label{sec:osc}

Experiments on ultracold Fermi gases are particularly suited for testing the strong coupling transport universality hypothesis because they offer the possibility of studying systems in two and three spatial dimensions, with tunable interaction strength, while offering direct real-time information about the density profile and correlations of the gas. The limit of strong interactions is achieved experimentally by tuning a bias magnetic field until the gas experiences a broad Feshbach resonance at which the s-wave scattering length $a$ diverges (unitary regime). Placing the gas of fermionic atoms in a deep optical trap with trapping frequencies $\omega_x,\omega_y,\omega_z$ in the x,y and z directions then allows experimentalists to routinely measure time-resolved oscillations of the radii of the gas cloud's shape, from which the frequency and damping of the underlying collective modes can be extracted \cite{Kinast:2005zz,Riedl:2008,Vogt:2011np}. The time evolution of the cloud's shape in two dimensions can be studied experimentally by making the trapping frequency $\omega_z$ much larger than both $\omega_{x},\omega_y$, so that the gas cloud is extremely compressed in the $z$-direction, and for $\omega_x\simeq \omega_y$ the cloud shape resembles that of a pancake. By contrast, if $\omega_z$ is much smaller than both $\omega_x,\omega_y$ the cloud shape resembles that of a cigar. Oscillations of the shape of the cigar in the x-y plane will be referred to as three dimensional dynamics in the following.

\begin{figure}[t]
\centering
\includegraphics[width=0.45\textwidth]{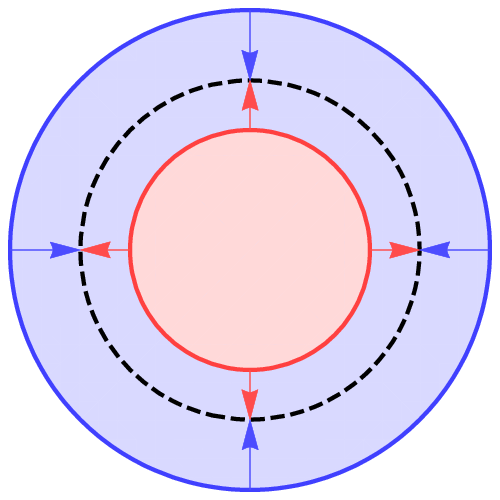}\hfill
\includegraphics[width=0.45\textwidth]{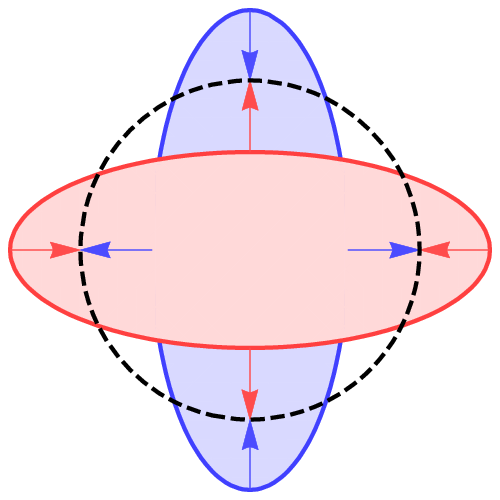}
\caption{\label{fig:one} Sketch of shape oscillations of an atomic gas cloud in the x-y plane:
breathing mode (left) and quadrupole mode (right). The breathing mode changes the overall cloud volume while the quadrupole mode corresponds to a surface deformation without volume change. The cloud's equilibrium configuration is indicated by the dashed circle.}
\end{figure}

Two independent cloud shape oscillation modes which will be discussed in the following are sketched in Fig.\ref{fig:one}. One distinguishes between a radial quadrupole mode, corresponding to elliptic deformations without volume change, and a breathing mode. The amplitude $Q(t)$ for the quadrupole mode can be accessed experimentally by measuring the difference in the width of the main axes, while the amplitude $B(t)$ for the breathing mode is obtained by summing the widths. 
A simple approximation to describing the time-evolution of the cloud's shape oscillations for a strongly interacting Fermi gas  is provided by the equations of hydrodynamics, more precisely the Navier-Stokes equations. (We will limit our discussion to temperatures above the superfluid phase transition where  single-fluid hydrodynamics is applicable). For a harmonic trapping potential and small oscillation amplitudes one finds analytic solutions to $B(t),Q(t)$ that are in the form of damped harmonic oscillations \cite{Schaefer:2009px,Brewer:2015hua}:
\begin{equation}
B(t)\propto\cos(\omega_B t) e^{- \Gamma_B t}\,,\quad Q(t)\propto\cos(\omega_Q t) e^{- \Gamma_Q t}\,.
\end{equation} 
 The frequencies $\omega$ and damping rates $\Gamma$ for these hydrodynamic modes differs between breathing and quadrupole mode. Specifically, one finds $\omega_{Q}=\sqrt{2}\,\omega_\perp$, $\Gamma_{Q}=\eta\, \omega_\perp^2/P$ for the hydrodynamic quadrupole mode in both $d=2$ and $d=3$ dimensions. In the case of the breathing mode, $\omega_{B}=\sqrt{10/3}\,\omega_\perp$, $\Gamma_{B}=\eta\, \omega_\perp^2/3 P$ for the $d=3$ breathing mode \cite{Brewer:2015ipa,Massignan:2005,Riedl:2008}. (The breathing mode in two dimensions is undamped and will not be considered in the following). In expressions above, $P$ denotes the local equilibrium pressure of the strongly interacting Fermi gas, $\omega_\perp\equiv \sqrt{\omega_x \omega_y}$ is the average trap frequency in the x-y plane and a constant ratio $\eta/P$ has been assumed in order to derive these analytic results.

\section{Black Hole Dual Calculation for Non-Hydrodynamic Frequencies and Damping Rates}
\label{sec:string}

The hydrodynamic modes should be contrasted with the corresponding collective modes expected from a string theory based approach. String theory based calculations employ so-called black hole duals to calculate properties of strongly interacting matter. In addition to hydrodynamic modes, black holes have an infinite number of non-hydrodynamic collective excitations (quasi-normal modes), which are similar to the ring-down modes of a glass struck (lightly) with a fork. As a working definition, non-hydrodynamic modes are modes that do not arise when studying solutions of the Navier-Stokes equations.
Despite recent progress \cite{Son:2008ye,Balasubramanian:2008dm,PhysRevD.85.106001}, no exact black hole dual description is known for a strongly interacting Fermi gas. However, one may attempt to use known duals that at least describe bulk features of cold atomic systems in $d$ (spatial) dimensions. In order to describe a non-relativistic strongly interacting liquid, we select a black hole dual that correctly reproduces the equation of state (the relation between pressure $P$ to energy density $\epsilon$) of a strongly interacting Fermi gas. Lifshitz black holes have a scaling parameter $z$ that enters the equation of state as $\epsilon z = P d$ \cite{Hoyos:2013eza}. Since a strongly interacting Fermi gas in $d$ dimension has $\epsilon = P d/2$ we choose $z=2$.

For Lifshitz black holes at finite temperature $T$ and zero chemical potential, which are the ones considered in the following, the ratio of shear viscosity to entropy density is known to be $\eta k_B/\hbar s=1/4\pi$ \cite{Taylor:2015glc}. In addition, the ring-down spectrum is straightforward to evaluate using a probe scalar with operator dimension $\Delta$. The operator dimension controls the type of perturbation considered. For instance for density perturbations (fermionic bilinears), the operator dimension would be $\Delta=d$, while for energy density perturbations (fermionic bilinears with a gradient), $\Delta=d+1$. While it is not known exactly which operator dimension corresponds to the case of density perturbations in the strongly interacting Fermi gas, a reasonable assumption is that $\Delta$ is bounded by the cases $\Delta=d$ and $\Delta=d+1$, so final results including systematic error estimates will be based on the mean and difference from these two choices. To be specific, we compute the quasinormal modes of a scalar field propagating in a fixed Lifshitz black brane background, using the setup described in \cite{Sybesma:2015oha}.

In the case of $d=z$, analytic expressions for the ring-down frequencies and damping rates were found in Ref. \cite{Sybesma:2015oha}, and we use these for $d=2$
\begin{equation}
\omega^{(d=2)}_n=0\,,\quad \Gamma_n^{(d=2)}= \left(n-1+\frac{\Delta}{2 z} \right)\times 4 \pi k_B T/\hbar\,,\quad n\geq 1\,.
\end{equation}
For $d=3$, we determine the corresponding values numerically following the general approach outlined in \cite{Starinets:2002br}, and collect the results in Table \ref{tab:one}. 

\begin{table}[t]
\centering
\begin{tabular}{|c|c|c|c|c|}
\hline
& \multicolumn{2}{c|}{$\Delta=3$} & \multicolumn{2}{c|}{$\Delta=4$}\\ \hline
$n$ & $\omega_n \hbar/(4\pi k_B T)$ & $\Gamma_n \hbar/(4\pi k_B T)$ & $\omega_n \hbar/(4\pi k_B T)$ & $\Gamma_n \hbar/(4\pi k_B T)$\\ \hline
1 & 0.2812 & 0.5282 & 0.3560 & 0.7540 \\ \hline
2 & 0.5776 & 1.437 & 0.6507 & 1.663 \\ \hline
3 & 0.8714  & 2.342 & 0.9446 & 2.568 \\ \hline
4 & 1.165  & 3.246 & 1.239 & 3.472 \\ \hline
\end{tabular}
\caption{\label{tab:one} Numerical results for frequencies and damping rates for ring-down frequencies for $d=3$ and two choices of $\Delta$. }
\end{table}

%\begin{table}[t]
%\centering
%\begin{tabular}{|c|c|c|c|c|}
%\hline
%{\bf n} & $\omega_n$ for $\Delta=3$ & $\Gamma_n$ for $\Delta=3$ & $\omega_n$ fo%r $\Delta=4$ & $\Gamma_n$ for $\Delta=4$\\
%\hline
%1 & $0.28123697 \times 4\pi k_B T /\hbar$ & $0.528167 \times 4\pi k_B T /\hbar$% & $0.3559963\times 4\pi k_B T /\hbar$  & $0.754037\times 4\pi k_B T /\hbar$ \\
%\hline
%\hline
%\end{tabular}
%\caption{Numerical results for frequencies and damping rates for ring-down freq%uencies for $d=3$ and two choices of $\Delta$. \textbf{Needs completion!}}
%\end{table}

The results in Table \ref{tab:one} all scale as $4 \pi T$ for a liquid of temperature $T$, vanishing chemical potential, and shear viscosity over entropy ratio of $\eta/s=\hbar/4 \pi k_B$. This situation differs from the case of real Fermi gases because no exact dual to real Fermi gases is known. Real Fermi gases have sizable chemical potential, small temperature and shear viscosity over entropy ratios different from $\eta/s=\hbar/4 \pi k_B$. In order to connect the string theory-based calculations to real Fermi gases, guidance from kinetic theory is employed. In kinetic theory, one encounters a single non-hydrodynamic mode with a damping rate $\Gamma_1=1/\tau_R$ where the relaxation time $\tau_R$ is known to obey $\tau_R\propto \eta/P$ in the hydrodynamic limit\cite{Brewer:2015ipa}. The results from Table \ref{tab:one} can be brought into this form under the assumption that $4\pi k_B T/\hbar\rightarrow s T/\eta$, which is trivially correct for the employed black hole dual. Using furthermore the thermodynamic identity $s T = (\epsilon+P)$ for a fluid with finite temperature and zero chemical potential, it is proposed that the replacement
\begin{equation}
\label{eq:myst}
4\pi k_B T/\hbar \rightarrow (\epsilon+P)/\eta\,, 
\end{equation}
in the results for the non-hydrodynamic frequencies and damping rates of Table \ref{tab:one}
can be used to connect the black hole dual calculations to real Fermi gases for strong interactions. Once this replacement has been performed, all explicit reference to temperature and chemical potential have been replaced by energy density $\epsilon$ and pressure $P$, which can be applied to a cold strongly interacting Fermi gas with equation of state $\epsilon=P d/2$ and different values of shear viscosity.

Finally, the above calculations are for the case of an untrapped Fermi gas, rather than a Fermi gas in an optical trap which is studied in most experimental setups. To relate the above untrapped results to the case of a trapping potential with average trapping frequency $\omega_\perp$, again guidance from kinetic theory is used. In kinetic theory, hydrodynamic mode oscillations change qualitatively between a free system and a system placed in a trap, but
%For instance, the hydrodynamic sound mode dispersion relation $\omega=c_s k-\frac{i \eta {\bf k}^2}{6 P}$, where $c_s$ is the local speed of sound and ${\bf k}$ is the wave number, gets modified to $\omega=\sqrt{10/3}\omega_\perp-\frac{i \eta \omega_\perp^2}{3 P}$ for the dispersion relation of the hydrodynamic breathing mode in three dimensions.
%However, in kinetic theory, 
the non-hydrodynamic modes 
%dispersion relation does 
do not change at all (cf. Ref.~\cite{Brewer:2015ipa}). Based on this observation, the non-hydrodynamic mode frequencies and damping rates from the black hole dual calculation of an untrapped system above are directly applied to the case of a trapped Fermi gas.

\subsection{Summary of Assumptions}

In making predictions for the properties of non-hydrodynamic modes in trapped unitary Fermi gases several assumptions have been made, which are summarized below:
\begin{itemize}
\item
Black holes in asymptotic Lifshitz spaces have been assumed to describe the bulk features of a strongly interacting Fermi gas
\item
A probe scalar with dimension $\Delta\simeq d+\frac{1}{2}$ has been assumed to approximately describe density perturbations in the strongly interacting Fermi gas
\item
It has been assumed that a strongly interacting Fermi gas at non-zero density and $\frac{\eta k_B}{s \hbar}\neq \frac{1}{4 \pi}$ is well approximated by the calculation for a Lifshitz black hole done at zero density and $\frac{\eta k_B}{s \hbar}= \frac{1}{4 \pi}$ when performing the \hbox {replacement (\ref{eq:myst})}
\item
It has been assumed that the frequencies and damping rates of non-hydrodynamic modes do not differ between the untrapped and trapped Fermi gas
\end{itemize}

All of these assumptions can in practice be tested and in most cases lifted by performing more general calculations. However, we leave these more demanding calculations for future work.

\section{Results}
\label{sec:res}

Let us consider density perturbations in a trapped, unitary Fermi gas. Besides the familiar hydrodynamic component, the dual black hole calculation implies that there are an infinite number of novel, non-hydrodynamic modes with relative amplitudes $\alpha_n$. A single non-hydrodynamic mode also is present in kinetic theory \cite{Brewer:2015ipa}; however, kinetic theory is a weak-coupling, particle-based description not quantitatively applicable to strongly interacting Fermi gases, so it is unclear how to interpret the kinetic theory result. If black hole duals can be used to describe real unitary Fermi gases then this 
implies that density perturbations give rise to generalized breathing and quadrupole modes of the form 
\begin{equation}
\label{eq:qnmform}
H(t)=\alpha_H\cos(\omega_H t+\phi_H) e^{- \Gamma_H t}+\sum_{n=1}^\infty \alpha_n \cos(\omega_n t+\phi_n) e^{- \Gamma_n t}\,,
\end{equation} 
where $H=B,Q$ depending on the shape oscillation considered, and possible phase shifts $\phi_H,\phi_n$ have been allowed. 

\begin{table}[t]
\centering
\begin{tabular}{|c|c|c|c|c|}
\hline
& \multicolumn{2}{c|}{$d=2$} & \multicolumn{2}{c|}{$d=3$}\\ \hline
$n$ & $\omega_n\times\eta/P$ & $\Gamma_n\times \eta/P$ & $\omega_n\times \eta/P$ & $\Gamma_n\times \eta/P$\\ \hline
%\hline
1 & 0 & $1.25(25) $ & $0.8(1) $ &  $1.6(3) $\\
\hline
2 & 0 & $3.25(25) $ & $1.5(1) $ &  $3.9(3) $\\
\hline
3 & 0 & $5.25(25) $ & $2.3(1) $ &  $6.1(3) $\\
\hline
4 & 0 & $7.25(25) $ & $3.0(1) $ &  $8.4(3) $\\
\hline
\end{tabular}
\caption{\label{tab:two} Numerical values for the frequencies and damping rates of the first $n\leq 4$ non-hydrodynamic modes in $d=2$ and $d=3$ dimensions, obtained from a string theory based calculation. Results are expressed in terms of the ratio of pressure $P$ to shear viscosity $\eta$. Note that $P/\eta$ can be re-expressed in terms of the  damping rates $\Gamma_Q,\Gamma_B$ of the hydrodynamic quadrupole and breathing modes for a strongly interacting Fermi gas in a trap.}
\end{table}

As outlined in section \ref{sec:string}, the frequencies  $\omega_n$ and damping rates $\Gamma_n$ of the non-hydrodynamic modes can be calculated with the black hole dual for both $d=2$ and $d=3$ dimensions. The results from section \ref{sec:string} have been condensed into experimentally accessible quantities for both $d=2,3$ shown in Table \ref{tab:two}. Note that for two dimensions, the result can be obtained analytically for all $n$, and one finds $\omega_n=0$ and $\Gamma_n=\left(2n-3/4\pm1/4\right)P/\eta$.

Note that, unlike the hydrodynamic component, the frequencies $\omega_n$ and damping rates $\Gamma_n$ of the non-hydrodynamic modes turn out to be independent of the cloud's average trapping frequency $\omega_\perp$. While the spatial oscillation structure of the non-hydrodynamic modes is exactly equal to those of the well-known hydrodynamic breathing and quadrupole modes, their respective time-dependent signature is quite different. Inspecting Table \ref{tab:two} it becomes apparent that the non-hydrodynamic modes are excitations with damping rates larger than the oscillation frequency in all cases. (For two dimensions, the analytic result implies that the excitations are purely damped, corresponding to an exponentially decreasing evolution without harmonic oscillations). The presence of an infinite number of non-hydrodynamic modes with specific frequencies and damping rates is a new prediction for strongly interacting Fermi gases. The different time-dependence offers an experimental handle to distinguish these novel non-hydrodynamic modes from well-studied hydrodynamic oscillations.

\begin{figure}
\centering
\includegraphics[width=0.45\textwidth]{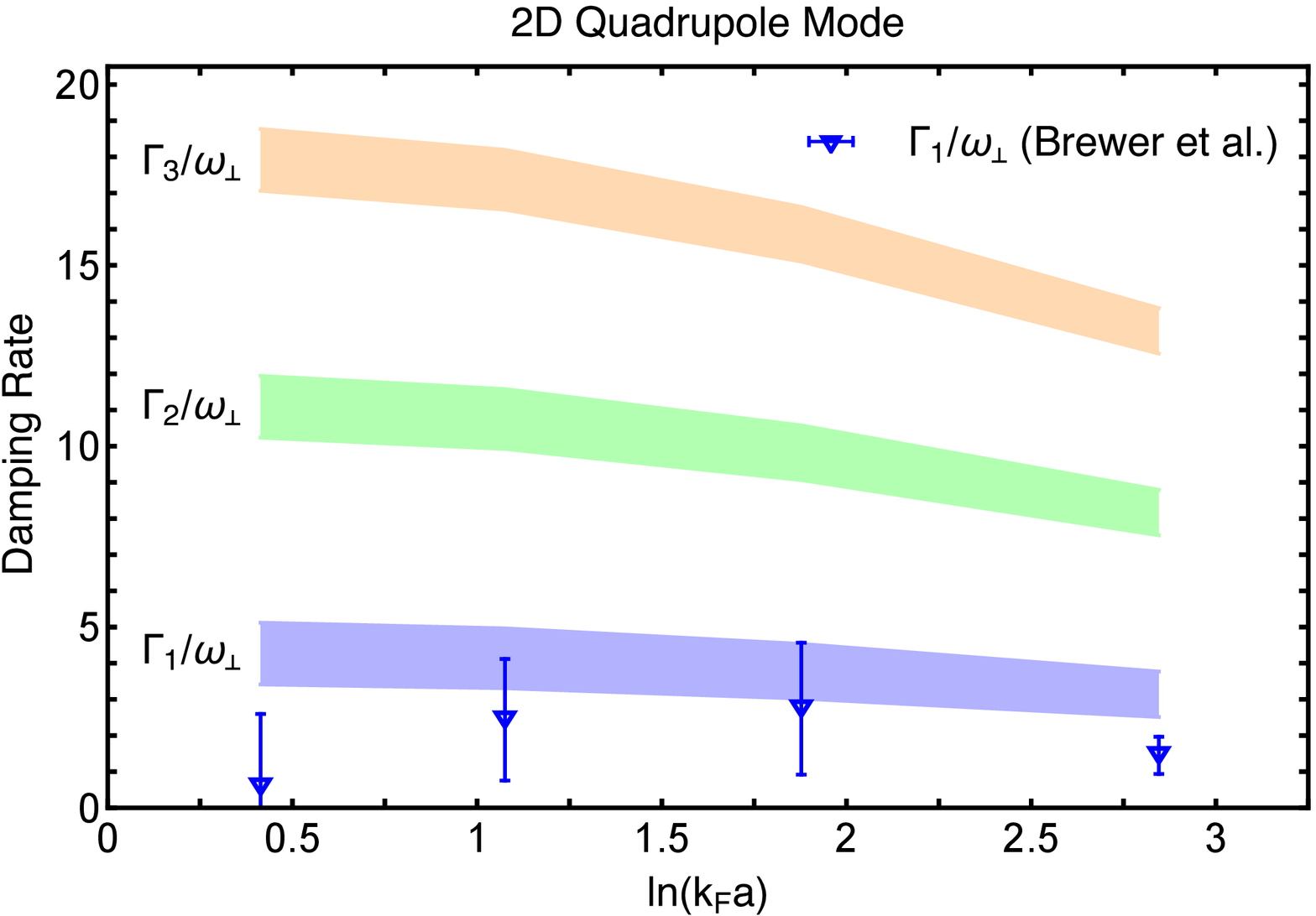}\hfill
\includegraphics[width=0.45\textwidth]{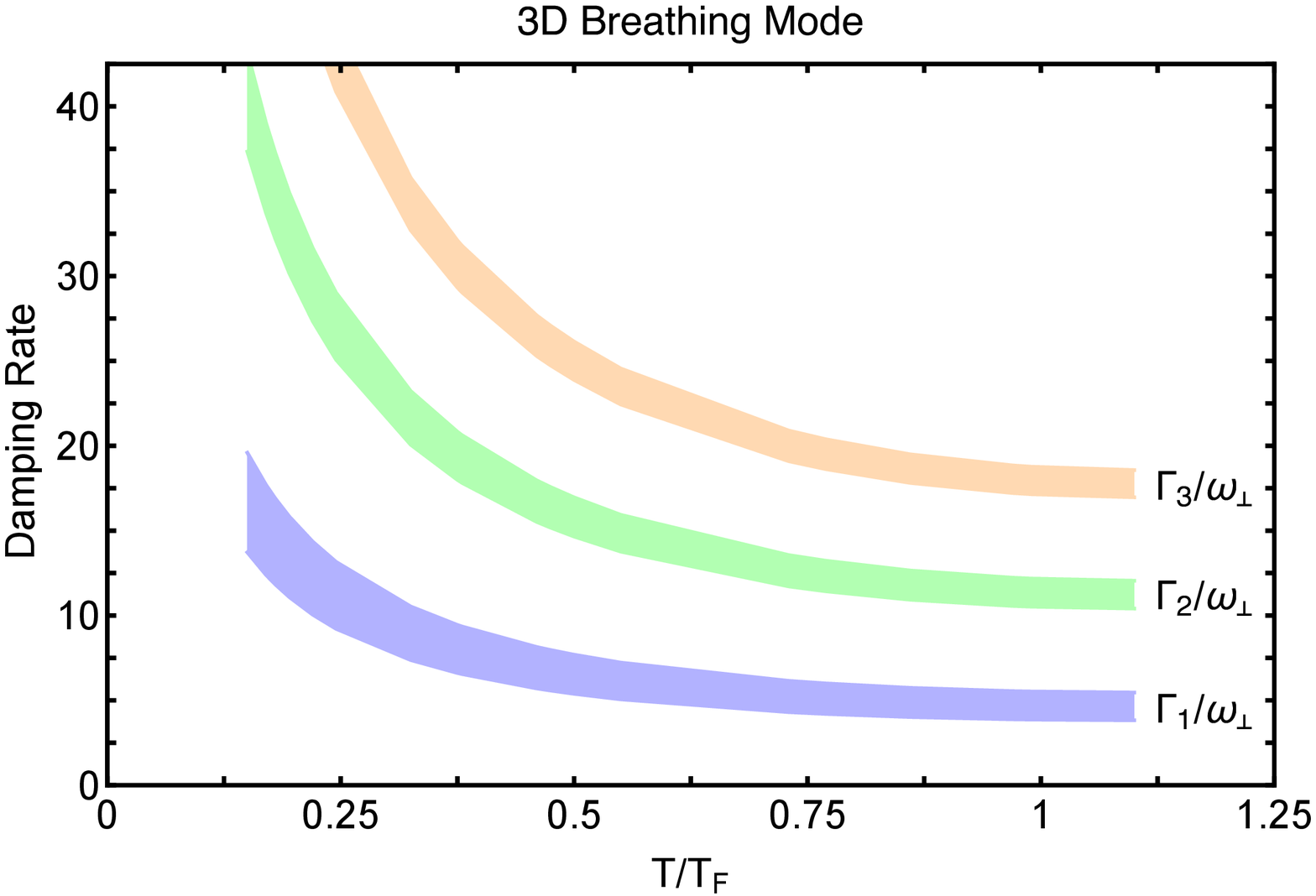}
\caption{\label{fig:two} Predicted non-hydrodynamic mode damping rates $\Gamma_{n}/\omega_\perp$ (bands).  Left: two-dimensional quadrupole mode as a function of interaction strength $\ln (k_F a)$, where $a$ is the two-dimensional s-wave scattering length and $k_F$ is the Fermi momentum \cite{abc}  related to the density $n$ at the cloud's center as $n=k_F^2/2\pi$. Also shown is the first non-hydrodynamic mode damping rate $\Gamma_1$ extracted from experimental data \cite{Brewer:2015ipa}.  Right: three-dimensional breathing mode as a function of cloud's temperature in units of the Fermi temperature $T_F=\left(3 N \omega_x \omega_y \omega_z\right)^{1/3}\hbar /k_B$ where $N\simeq 5\times 10^5$ is the number of atoms in the three-dimensional optical trap.} %See text for details.}
\end{figure}

In order to facilitate experimental detection of these non-hydrodynamic collective modes, results from Table \ref{tab:two} have been converted into predictions in Fig.~\ref{fig:two} for damping rates expressed in terms of the experimentally measured quantities. For this conversion, the relation between the hydrodynamic damping rates and $\eta/P$ discussed in section \ref{sec:osc} has been used to re-express the predicted non-hydrodynamic damping rates from Table \ref{tab:two} in terms of the damping rates $\Gamma_Q,\Gamma_B$ for the hydrodynamic modes\cite{exp2}. The damping rates $\Gamma_Q,\Gamma_B$ themselves have been measured experimentally in two and three dimensions, respectively \cite{Kinast:2005zz,Vogt:2011np,Riedl:2008}. For example, $\Gamma_1=1.25(25) \frac{P}{\eta}$ for the $d=2$ quadrupole mode from Table \ref{tab:two} becomes $\Gamma_1=1.25(25) \Gamma_Q^{-1} \omega_\perp^{2}$ which becomes $\Gamma_1\simeq 4.17(83)\omega_\perp$ when using the experimentally determined value of $\Gamma_Q\simeq 0.30\omega_\perp$ from Ref.~\cite{Vogt:2011np} at $\log(k_F a)\simeq 1.04$.  
 In Fig.~\ref{fig:two}, predictions for non-hydrodynamic mode damping rates $\Gamma_n/\omega_\perp$ are shown in the case of the two-dimensional quadrupole mode  and the three dimensional breathing mode, given different choices of temperature and atom interaction strength. For reference, also shown in Fig.~\ref{fig:two} are 
%damping rates extracted from experimental data for the hydrodynamic modes $\Gamma_Q,\Gamma_B$ \cite{Kinast:2005zz,Vogt:2011np,Riedl:2008} as well as 
the only published constraints on the two-dimensional quadrupole mode damping rate $\Gamma_1$ extracted from experimental data\cite{Brewer:2015ipa}. 
In Ref.~\cite{Brewer:2015ipa}, existing experimental data \cite{Vogt:2011np} on the quadrupole mode in a two-dimensional trapped Fermi gas had been re-analyzed using a two-component form (\ref{eq:qnmform}). Given the time-resolution and number of data-sets obtained in the experiment, the re-analysis performed in Ref.~\cite{Brewer:2015ipa} found that information about a predicted second component could be extracted from the data, albeit with low statistical significance. This can be understood through the fact that the measured amplitude $Q(t)$ is most sensitive to the predicted non-hydrodynamic component in Eq.~(\ref{eq:qnmform}) at early times, thus requiring a high time-sampling frequency. In contrast, the main aim in Ref.~\cite{Vogt:2011np} had been extraction of the hydrodynamic component which requires long-time information. As a consequence, the time-sampling rate chosen in Ref.~\cite{Vogt:2011np} is not optimal to extract early-time information with high statistical significance.  It is likely that the statistical significance of the non-hydrodynamic component could be vastly improved if the experiment in Ref.~\cite{Vogt:2011np} could be repeated to yield 100 data sets with time-resolution increased by a factor of 20. Absent newer experimental data, it is nevertheless interesting to note that the experimental constraints on $\Gamma_1$ obtained in Ref.~\cite{Brewer:2015ipa} are broadly consistent with the present predictions.

It is also possible to estimate the amplitude ratio $\alpha_1/\alpha_H$ of the first non-hydrodynamic mode relative to the amplitude of the usual hydrodynamic mode when assuming that standard experimental procedures are used to excite quadrupole and breathing mode oscillations \cite{Kinast:2005zz,Vogt:2011np,Riedl:2008}. Assuming random phase shifts $\phi_H,\phi_1$ for individual hydrodynamic and non-hydrodynamic components, one finds 
$|\frac{\alpha_1}{\alpha_H}|\simeq (\omega_H+\Gamma_H)/(\omega_1+\Gamma_1)$. This result would imply typical amplitude ratios of $|\frac{\alpha_1}{\alpha_H}|\simeq 40$ percent for the two-dimensional quadrupole mode and $|\frac{\alpha_1}{\alpha_H}|\simeq 20$ percent for the three-dimensional breathing mode. This estimate places the predicted first non-hydrodynamic mode well within the experimental detection capabilities of present state-of-the-art experiments in both two and three dimensions.

\section{Summary and Conclusions}
\label{sec:conc}

In this work, strong coupling calculations based on string theory have been used to make quantitative predictions for the existence and properties of novel, non-hydrodynamic modes in strongly interacting Fermi gases. The temporal signatures, frequencies, damping rates as well as amplitude of the first non-hydrodynamic mode relative to the well-measured hydrodynamic mode have been predicted for both two and three spatial dimensions. These predictions should be well within the experimental testing capabilities for current state-of-the-art experiments.

Indeed, in section \ref{sec:res} it was found that experimental constraints on the non-hydrodynamic mode damping rate $\Gamma_1$ in the two-dimensional case are consistent with the present string-theory based prediction \cite{Brewer:2015ipa}. 
However, given the large error bars and weak statistical significance of $\Gamma_1$ in the analyzed data, this is not sufficient to confirm our present predictions. New, high-precision experimental data would be required to confirm (or rule out) the presence of these novel, non-hydrodynamic modes, and thus provide a first stringent test of strong coupling transport universality.

\acknowledgements
This work was supported in part by the Department of Energy, DOE award No. DE-SC0008132, grant Contract Number DE-SC0011090 and European Research Council Starting Grant grant no. NewNGR-639022. We would like to thank T.~Enss, U.~Romatschke, J.~Thomas, T.~Sch\"afer and W.~van der Schee for fruitful discussions on this topic.


\begin{thebibliography}{99}
\bibitem{abc}
Unlike the case of three dimensions, the scattering amplitude in two dimensions depends logarithmically on $k_F a$, which makes $\ln (k_F a)$ a good interaction parameter similar to the role played by $1/(k_F a)$ in three dimensions \cite{bloch}.
\bibitem{exp2}
Having expressed the non-hydrodynamic mode damping rates in terms of $\Gamma_Q,\Gamma_B$ allows predictions for variable (but large) interaction parameter even though the underlying calculation is only strictly valid at infinite interaction strength (see Methods section for details).
\end{thebibliography}
\end{document}